\documentclass[aps,prd,preprint,floats,epsf,superscriptaddress,nofootinbib]{revtex4-1}

\usepackage{graphicx} 
\usepackage{color}
\usepackage{xcolor}
\usepackage{hyperref}
\usepackage{amssymb}
\usepackage{booktabs}
\usepackage{rotating}
\usepackage{amsmath}
\usepackage{multirow}
\usepackage{enumitem}

\providecommand{\aj}{AJ}
\providecommand{\apj}{ApJ}
\providecommand{\apjl}{ApJL}

\providecommand{\aap}{A\&A}

\providecommand{\mnras}{MNRAS}

\providecommand{\prd}{Phys.~Rev.~D}

\providecommand{\jcap}{JCAP}

\newcommand{\comment}[1]{} 

\makeatother
\begin{document}
\title{A Universal Distribution of Dark Matter in Milky Way-like galaxies and How to Infer It}

\author{Sam Cheng-Tse Huang}
\email[e-mail: ]{sam.huang@physics.rutgers.edu}
\affiliation{NHETC, Department of Physics and Astronomy, Rutgers, Piscataway, NJ 08854, USA}

\author{Matthew R.~Buckley}
\email[e-mail: ]{mbuckley@physics.rutgers.edu}
\affiliation{NHETC, Department of Physics and Astronomy, Rutgers, Piscataway, NJ 08854, USA}

\author{Justin I. Read}
\affiliation{University of Surrey, Physics Department, Guildford, GU2 7XH, UK}

\author{David Shih}
\affiliation{NHETC, Department of Physics and Astronomy, Rutgers, Piscataway, NJ 08854, USA}

\date{\today}

\begin{abstract}
The phase-space density of dark matter within the Milky Way is a key quantity that encodes information about the nature of the dark sector. The local phase-space density is also required to properly interpret the results of dark matter direct detection experiments. However, there are at present few observational constraints. 
In this paper, we show that a simple coordinate transformation reveals a near-universal DM phase-space distribution function among three independent suites of cosmological simulations of Milky Way-mass galaxies. We provide evidence for this with plots of kinematic features as well as  machine learning-based classifiers that are sensitive to all of the correlations in the full multivariate phase space. 
Deviations from universality are found only at extremes of galactic radius and/or velocity, and in one simulation that has a prominent accreted dark disc. We further show that the parameters for the coordinate transformation can be inferred from metal poor stars ($\log_{10}[{\rm Fe}/{\rm H}]< -2$). These stars also contain signatures of the dark disk, allowing the existence of such a structure to be inferred from observation. Finally, we construct a model of this universal phase-space distribution using a normalizing flow, trained on the standardized phase-space across simulations. We will apply our method to survey data from Gaia and SDSS in a forthcoming work.
\end{abstract}

\maketitle
\section{Introduction}

Multiple observational probes across a wide range of astronomical and cosmological scales \citep{Zwicky:1933gu, Rubin:1980zd, Bertone:2016nfn,Planck:2018vyg,DESI:2024mwx,Cirelli:2024ssz,Abdalla:2022yfr} demonstrate that 25\% of the Universe \cite{WMAP:2003ivt,WMAP:2003pyh,Planck:2018vyg} is composed of dark matter (DM), a new form of stress-energy that was kinematically cold in the early Universe (CDM) \citep{ReadErkal19,nadler21,keeley24,irsic17,villasenor23,banik21}. Gravitational lensing of merging galaxy clusters \citep[e.g.,][]{Clowe2006,harvey15} and an apparent anti-correlation between central DM density and stellar mass in dwarf galaxies \citep{Read19} suggest that DM acts, to a good approximation, as a collisionless fluid. So far DM has escaped experimental detection in Earth-based facilities \citep{Lewin:1995rx,Freese:2012xd}, production in colliders \citep{ParticleDataGroup:2024cfk, Ilten:2022lfq}, and searches for its decay or annihilation byproducts in space \citep{Planck:2018vyg,Safdi:2022xkm,Cirelli:2024ssz}.

Understanding the properties of gravitationally-bound collections of DM across all scales -- from the cosmological down to dwarf galaxies and smaller -- allows for tests of the CDM paradigm. One such test is inferring the DM velocity distribution function (VPDF) and comparing this with model predictions. The DM VPDF encodes information about DM self-interactions \citep[e.g.,][]{Cirelli:2024ssz,Rocha:2012jg,Peter:2012jh, Tulin:2017ara, Spergel:1999mh,vandenbosch26} and whether DM is wave-like \citep[e.g.,][]{schive14}. Applied to our own galaxy, the Milky Way (MW), it is also of  critical importance as input to the direct and indirect detection experimental efforts~\citep{Drukier:1984vhf,Goodman:1984dc,Drukier:1986tm,Freese:2012xd,Necib:2018igl,Read:2014qva}.

 A longstanding paradigm for the DM VPDF is the Standard Halo Model (SHM)~\citep{1986PhRvD..33.3495D, 1996APh.....6...87L}. Motivated by early simulations in CDM, the SHM approximates the phase-space density of a galaxy's DM halo as a smooth, static system with a simple analytic density radial profile (such as an isothermal, Navarro-Frenk White (NFW), or Einasto~\citep{Navarro:1996gj} profile), and a Maxwellian VPDF~\citep{Drukier:1986tm} with a radius-dependent characteristic velocity. This characteristic velocity requires knowledge of the total enclosed mass, which can introduce $\sim 30\%$ uncertainties at the Solar radius \citep{Posti_2019,Watkins_2019,2022ApJ...925....1S,2024ApJ...972...70R}. Additionally, this framework neglects baryonic feedback~\citep{Sloane:2016kyi}, satellite accretion~\citep{Read:2008fh}, and disk evolution~\citep{debattista08,Ling_2010}, including heating by a stellar bar \citep[e.g.,][]{pedro06}. These astrophysical effects result in deviations from the SHM, and can modify the DM phase-space density enough to significantly alter expected DM particle detection signals~\citep{freese04,bruch09,Sloane:2016kyi}.

More recent efforts have attempted to go beyond the SHM by leveraging detailed aspects of cosmological simulations to make connections with observations. \cite{tissera14} used cosmological simulations to show that low metallicity stars ($\log_{10}[{\rm Fe}/{\rm H}] < -3$) trace the underlying DM density and so could be used to constrain DM halo shape. \cite{Herzog-Arbeitman:2017fte,herzog-arbeitman18} explore a similar idea, but using low metallicity stars to probe the VPDF of DM. They argue that higher metallicity stars ($\log_{10}[{\rm Fe}/{\rm H}] <-1.5$), which are more accessible observationally, can also be used to constrain the DM VPDF. \cite{ruchti14,ruchti15} were the first to suggest that significant accreted DM components should come with accreted stars with distinct chemistry. They found no evidence for such stars in the Gaia-ESO survey data. They thus argued that the MW is unlikely to have a significant accreted disc of DM, and therefore its local VPDF should be close to the canonical SHM model. More recent work using Gaia data and associated surveys have found complex chemo-dynamic substructure in the disc \citep[e.g.,][]{necib20,wang23,sestito26}, but it is not clear whether these correspond to accreted dwarfs and hence have associated accreted DM. Finally, \citep{Necib:2018igl,shpigel26} use numerical simulations of MW-mass galaxies to show that accreted stars, decomposed into their separate merger origins, can be used to constrain the local VPDF of the MW.

In this paper, we build on the above works by using both position and velocity information of low metallicity (i.e., accreted) stars to construct a statistical map between stars and the co-local DM, avoiding reliance on measurements of the enclosed mass. 
To ensure a robust map, we use three independent suites of cosmological simulations of MW-mass galaxies to test and refine our methodology: \textbf{N-body Shop}~\citep{Stinson_2012,Zolotov_2012,Loebman_2012}, \textbf{VINTERGATAN-GM}~\citep{Agertz2021,2022MNRAS.510.4208R,Joshi:2023mht,Rey:2022mwh,Rodriguez-Cardoso:2026lty}, and \textbf{FIRE-2}~\citep{Hopkins:2017ycn,Wetzel:2016wro, Garrison_Kimmel_2017,Debattista_2019}. We focus on simulated stars and DM close enough to the equivalent of the Solar radius that the stellar kinematics can be reconstructed by the {\it Gaia} Space Telescope, allowing our method to be applied to data in future work.

We first show that a simple rescaling of position and velocity transforms the DM phase-space density distributions from all of these simulations onto a single universal DM profile. This improves on prior work in the literature that rescaled positions only \citep{Necib:2018igl, Folsom:2025lly}. 
In particular, \citep{Folsom:2025lly}'s energy-conserving rescaling collapses the 1D DM speed distribution onto a near-Maxwellian form, which we adopt in the same spirit as our baseline, then go further by standardizing the full 6D phase-space. 
We then use machine learning classifier algorithms to show that this rescaling outperforms other methods proposed in the literature, allowing us to accurately predict the local VPDF of DM of each simulation (with a mean error of 10\%), when training on the others. While variations between simulation suites and individual galaxies remain, these differences are mostly concentrated at large and small galactic radii: in moderate distances corresponding to the approximate Solar location, the phase-space distributions can be scaled to be remarkably uniform. We go on to demonstrate that this global scaling can be inferred directly from local metal-poor stellar tracers~\citep[c.f.,][]{tissera14,Herzog-Arbeitman:2017fte,herzog-arbeitman18,Necib:2018iwb}, enabling the recovery of robust DM distributions using only observable quantities. 

The method recovers the DM VPDF at an accuracy of 10\% near the Solar neighbourhood across all simulations, with the exception of one --- the FIRE-2 simulation \texttt{m12m} --- that has a prominent accreted dark disc \citep[]{Read:2008fh}. In this case, our recovery degrades to 25\%. However, the metal poor stars in this simulation show clear rotation, with amplitude 150\,km/s at the Solar position. This is readily detectable in data and so can be used to test or constrain the presence of a dark disc in the MW, and to determine whether our universal scaling method to recover the DM distribution can be robustly applied. We will apply our new method to real data to determine the VPDF in the MW in a companion paper.

Finally, using normalizing flows~\citep{rezende2016variationalinferencenormalizingflows}, we create a generative model of the DM phase-space density which can be inferred from the distribution of low-metallicity stars within a galaxy. Our universal relation serves as a preprocessing step in the flow training, enabling high-fidelity reconstruction of halo kinematics from pooled multi-simulation datasets. When applied to real data, this flow can be used to generate a data-driven numerical model of the local velocity distribution of DM.

This paper is organized as follows. Section~\ref{sec:simulations} details the simulation suites and preprocessing. Section~\ref{sec:classifier} defines the unified scaling strategy, while Section \ref{sec:BDT} quantifies DM universality via classifier-based inference. Section~\ref{sec:flows} demonstrates the generative efficacy of the scaling strategy using normalizing flows. We conclude in Section~\ref{sec:conclusion}. 

\section{Data} \label{sec:simulations}
\subsection{Simulated Galaxies}

\begin{figure}[tp]
    \centering
    \includegraphics[width=1.\linewidth]{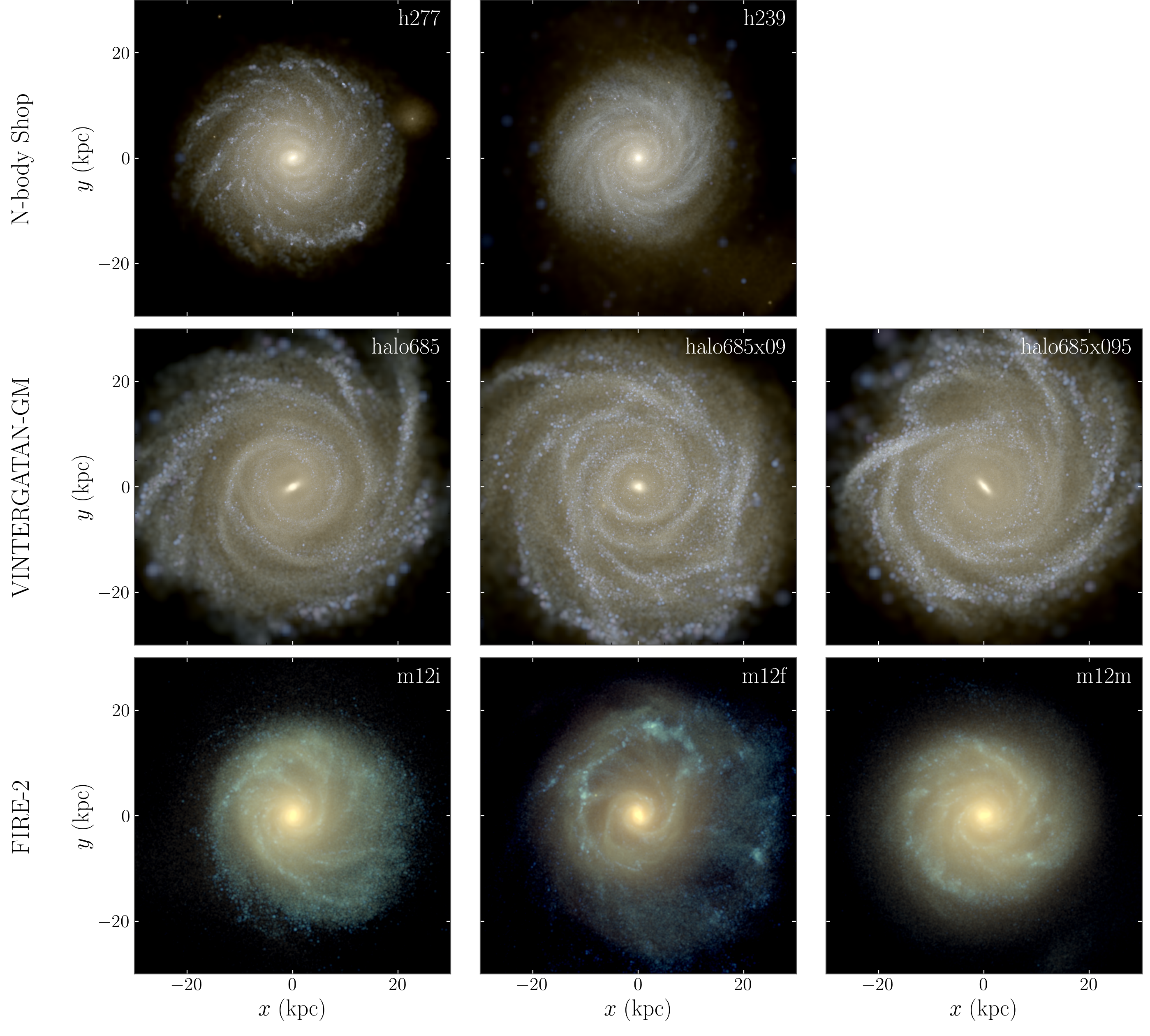}
\caption{Stellar part of the simulated Milky Way-like galaxies at redshift $\mathrm{z}=0$ for \textbf{N-body Shop}, \textbf{VINTERGATAN-GM} and \textbf{FIRE-2} within 30~kpc.}
    \label{fig:MW 2D}
\end{figure}

We analyse eight Milky Way-mass galaxies at redshift $z=0$ drawn from three high-resolution cosmological zoom-in simulation suites: \textbf{N-body Shop}, \textbf{VINTERGATAN-GM}, and \textbf{FIRE-2}. Face-on views of the baryonic disks of the galaxies are shown in Figure~\ref{fig:MW 2D}. All three suites share a standard $\Lambda$CDM cosmological framework ($H_0 \approx 70~\mathrm{km~s^{-1}~Mpc^{-1}}$, $\Omega_m \approx 0.3$, $\Omega_\Lambda \approx 0.7$) and utilize a ``zoom-in'' technique to focus on Milky Way-mass galaxies at high resolution within a consistent large-scale environment. The mass resolution in stars and DM for each simulation is given in Table \ref{tab:halo_details}. However, each simulation suite employs distinct hydrodynamical solvers and subgrid feedback prescriptions, yielding phenomenological differences in the resulting stellar and DM distributions. We briefly summarise these below.

\textbf{N-body Shop} uses the Smoothed
Particle Hydrodynamics (SPH) \citep{1977AJ.....82.1013L, 1977MNRAS.181..375G} code \textsc{Gasoline}~\citep{Wadsley:2003vm} with a ``blastwave'' feedback model \citep{Stinson:2006cp}. This formalism temporarily disables radiative cooling ($\sim 10~\mathrm{Myr}$) to mimic adiabatic supernova expansion; such feedback can impulsively heat the central potential, potentially transforming primordial DM cusps into constant-density cores \citep{Pontzen:2011ty, Di_Cintio_2013}.

\textbf{VINTERGATAN-GM} employs \textsc{Ramses} \citep{Teyssier:2001cp}, an Adaptive Mesh Refinement (AMR) \citep{1989JCoPh..82...64B} code  with explicit thermal and kinetic feedback. Unlike the blastwave method, this implementation typically results in the adiabatic contraction of the DM halo in response to baryon accumulation, preserving the central density cusp \citep{Agertz2021}. Notably, its genetically modified extension alters the initial conditions to control specific merger mass ratios while fixing the final virial mass.

\textbf{FIRE-2} utilizes the Meshless Finite-Mass (MFM) solver in \textsc{Gizmo}~\citep{Hopkins:2014qka} with multi-channel explicit feedback (mechanical, radiative, cosmic rays) directly from stellar population models \citep{Hopkins:2017ycn}. This approach generates self-consistent ``bursty'' star formation without cooling shutoffs. The resulting fluctuations remain strong at the Milky Way scale and drive a departure from adiabatic contraction. \citep{Wetzel_2016, Lazar_2020}.

In addition to kinematic variables, the simulated star particles (each representing a collection of individual stars) are labeled with stellar metallicity $\log_{10}[{\rm Fe/H}]$. Low metallicity stars are typically old, and for Milky Way-type galaxies, mostly formed outside the main halo. We therefore use metallicity to identify such ``ex-situ'' stars, compared to the high-metallicity stars that formed in-situ in the galactic disks~\citep{Herzog-Arbeitman:2017fte, Necib:2018iwb}.

The selected halos from the simulation suites have a diverse range of dynamical histories, as summarized in Table~\ref{tab:halo_details}. The \textbf{N-body Shop} halos bracket the extrema of merger activity: \texttt{h277}~\citep{brooks_2020_0653h-as743} is a quiescent Milky Way analogue (no major mergers at redshift $z<3$) favoring a stable disk and
oblate halo, while \texttt{h239} is a bulge-dominated remnant of active accretion. The \textbf{VINTERGATAN-GM} subset allows for a controlled study of merger impact, comprising a reference halo with a massive Gaia-Sausage/Enceladus-like (GSE-like) \citep{Belokurov_2018,Helmi_2018} merger (redshift $z \approx 1.5$) and two variants where this merger's mass ratio is systematically suppressed. This controlled variance isolates the role of massive accretion events in driving phase-space mixing and inducing halo triaxiality \citep{Rey:2021ecx}. The \textbf{FIRE-2} halos capture specific redshift $z=0$ dynamical features, including a standard isolated disk (\texttt{m12i}), a system perturbed by a massive Large Magellanic Cloud-like satellite (\texttt{m12f}), and a late-forming galaxy exhibiting a strong central bar (\texttt{m12m})~\citep{Debattista_2019}. 

To ensure a consistent dynamical baseline across suites with varying native virial definitions, we standardize our analysis using the virial radius $R_{\mathrm{vir}}$, defined as the radius enclosing a mean density of $178 \times \rho_{\mathrm{crit}}$~\citep{Cole:1995ep}. 
We process the snapshots using the \textsc{Pynbody} framework \citep{2013ascl.soft05002P} for N-body Shop and Vintergatan halos and \textsc{Gizmo} framework 
for FIRE-2 halos. For all halos, we define the ``galactocentric'' coordinate system by identifying the halo baryocenter and orienting the system such that the stellar disk lies in the $x-y$ plane, with net rotation proceeding in the positive $y$-direction. 

\begin{table*}[tp]
\centering
\renewcommand{\arraystretch}{1.}
\setlength{\tabcolsep}{4pt}
\footnotesize
\begin{tabular}{l c c c c c c c p{4.0cm}}
\hline\hline
\textbf{Halo ID} & \textbf{$M_{\mathrm{vir}}$} & \textbf{$R_\mathrm{vir}$} & \textbf{$m_{\mathrm{DM}}$} & \textbf{$m_{\star}$} & \textbf{$m_{\mathrm{gas}}$} & \textbf{$\epsilon$} & \textbf{$\Delta x$} & \textbf{History / Morphology} \\
 & $[10^{12} M_{\odot}]$ & $[\mathrm{kpc}]$ & $[10^5 M_{\odot}]$ & $[10^4 M_{\odot}]$ & $[10^4 M_{\odot}]$ & $[\mathrm{pc}]$ & $[\mathrm{pc}]$ & \\
\hline
\multicolumn{9}{l}{\textbf{N-body Shop}~\citep{Stinson_2012,Zolotov_2012,Loebman_2012}} \\
\texttt{h277} & $0.69$ & 172 & $1.3$ & $0.58$ & $2.7$ & $187$ & $19$ & Quiescent; no major mergers $z<3$ \\
\texttt{h239} & $0.91$ & 186 & $1.3$ & $0.58$ & $2.7$ & $187$ & $19$ & Active; multiple mergers $z<1$; Bulge \\
\hline
\multicolumn{9}{l}{\textbf{VINTERGATAN-GM~\citep{Agertz2021,Joshi:2023mht,Rey:2022mwh,Rodriguez-Cardoso:2026lty}}} \\
\texttt{halo685} & $1.32$ & 203 & $2.0$ & $1.0$ & $3.6$ & $20$ & $20$ & massive GSE-like merger \\
\texttt{halo685x09} & 1.30 & 204 & $2.0$ & $1.0$ & $3.6$ & $20$ & $20$ & Merger mass ratio reduced to 0.9 \\
\texttt{halo685x095} & 1.30 & 204 & $2.0$ & $1.0$ & $3.6$ & $20$ & $20$ & Merger mass ratio reduced to 0.95 \\
\hline
\multicolumn{9}{l}{\textbf{FIRE-2}~\citep{Hopkins:2017ycn,Wetzel:2016wro, Garrison_Kimmel_2017,Debattista_2019}} \\
\texttt{m12i} & $1.18$ & 203 & $0.35$ & $0.7$ & $0.71$ & $4\,/\,40$ & $1$ & Isolated; typical disk \\
\texttt{m12f} & $1.64$ & 226 & $0.35$ & $0.7$ & $0.71$ & $4\,/\,40$ & $1$ & Active; LMC-mass companion \\
\texttt{m12m} & $1.50$ & 224 & $0.35$ & $0.7$ & $0.71$ & $4\,/\,40$ & $1$ & Late-forming; Strong bar \\
\hline\hline
\end{tabular}
\caption{Total halo mass~$M_{\mathrm{vir}}$, virial radius~$R_{\mathrm{vir}}$, and simulation mass resolution of the dark-matter, stellar and gas elements~$m_{\mathrm{DM}}$, $m_{\star}$ and $m_{\mathrm{gas}}$ of selected simulated halos at redshift $\mathrm{z}=0$. $R_{\mathrm{vir}}$ is defined by densities $\rho(R_{\mathrm{vir}}) \equiv 178\times\rho_{\mathrm{crit}}$, where $\rho_\mathrm{crit}$ is the critical density. $m_{\star}$ is the initial stellar particle mass and $m_{\mathrm{gas}}$ the initial gas resolution (gas-particle mass for the SPH/MFM runs, minimum cell mass for the AMR runs). $\epsilon$ is the gravitational (Plummer-/spline-equivalent) force softening length; two values denote star\,/\,DM particles where these differ. $\Delta x$ is the hydrodynamic spatial resolution: the (adaptive) SPH smoothing length for the GASOLINE/{\sc ChaNGa} and {\sc Gizmo}-MFM runs and the minimum AMR cell size for the {\sc Ramses} runs. For the Lagrangian (SPH/MFM) suites $m_{\mathrm{gas}}$ and the smoothing length are comparable to the baryon mass unit and fall below $\epsilon$ in dense gas; for FIRE-2 the gas softening/smoothing is adaptive with a $1\,$pc floor, rising to $\sim20$--$40\,$pc in the typical ISM.
}
\label{tab:halo_details}
\end{table*}

\begin{figure}[htp]
    \centering
    \includegraphics[width=0.9\columnwidth]{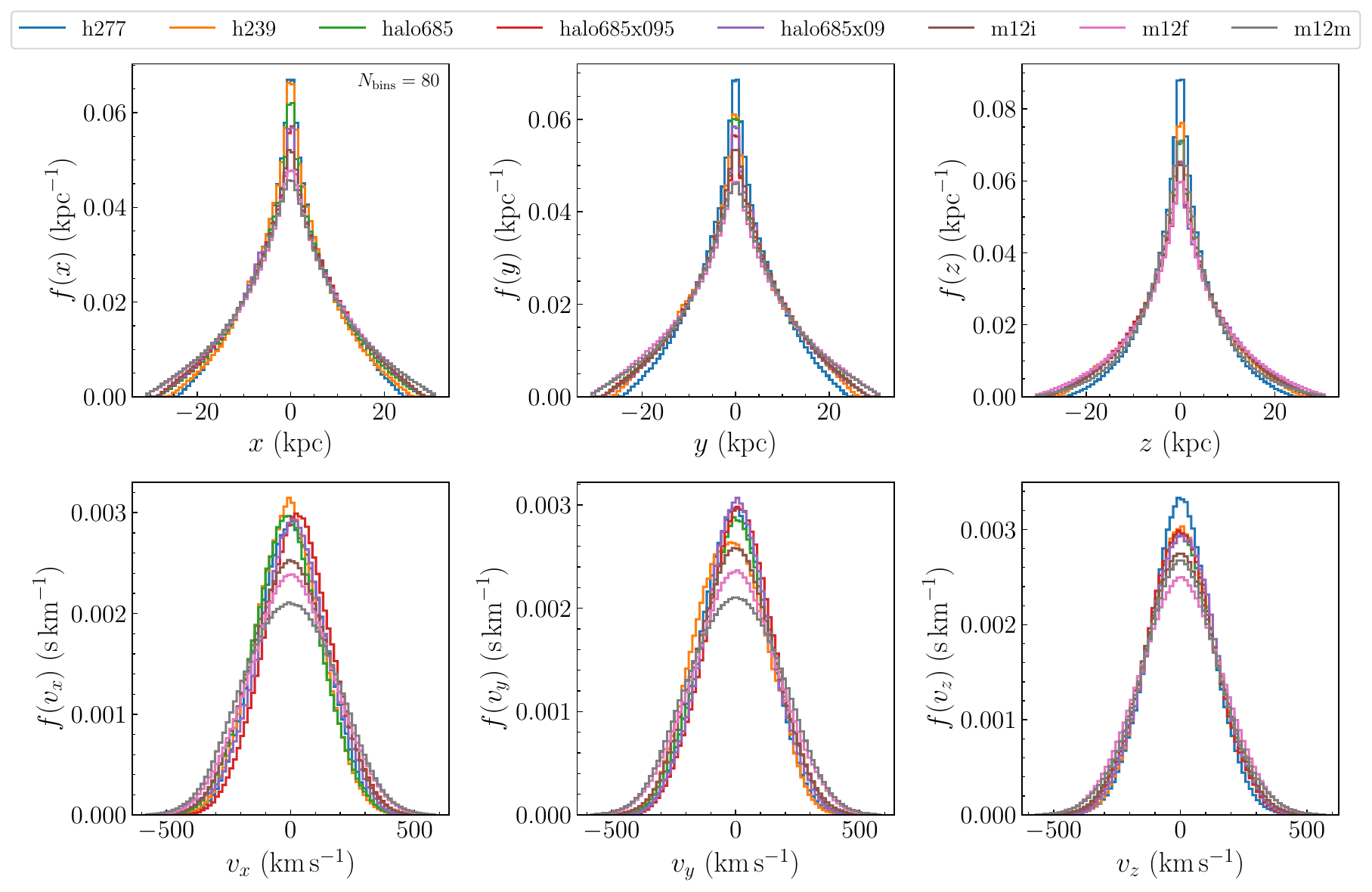}\label{fig:hist1d raw DM cart}
    \includegraphics[width=0.9\columnwidth]{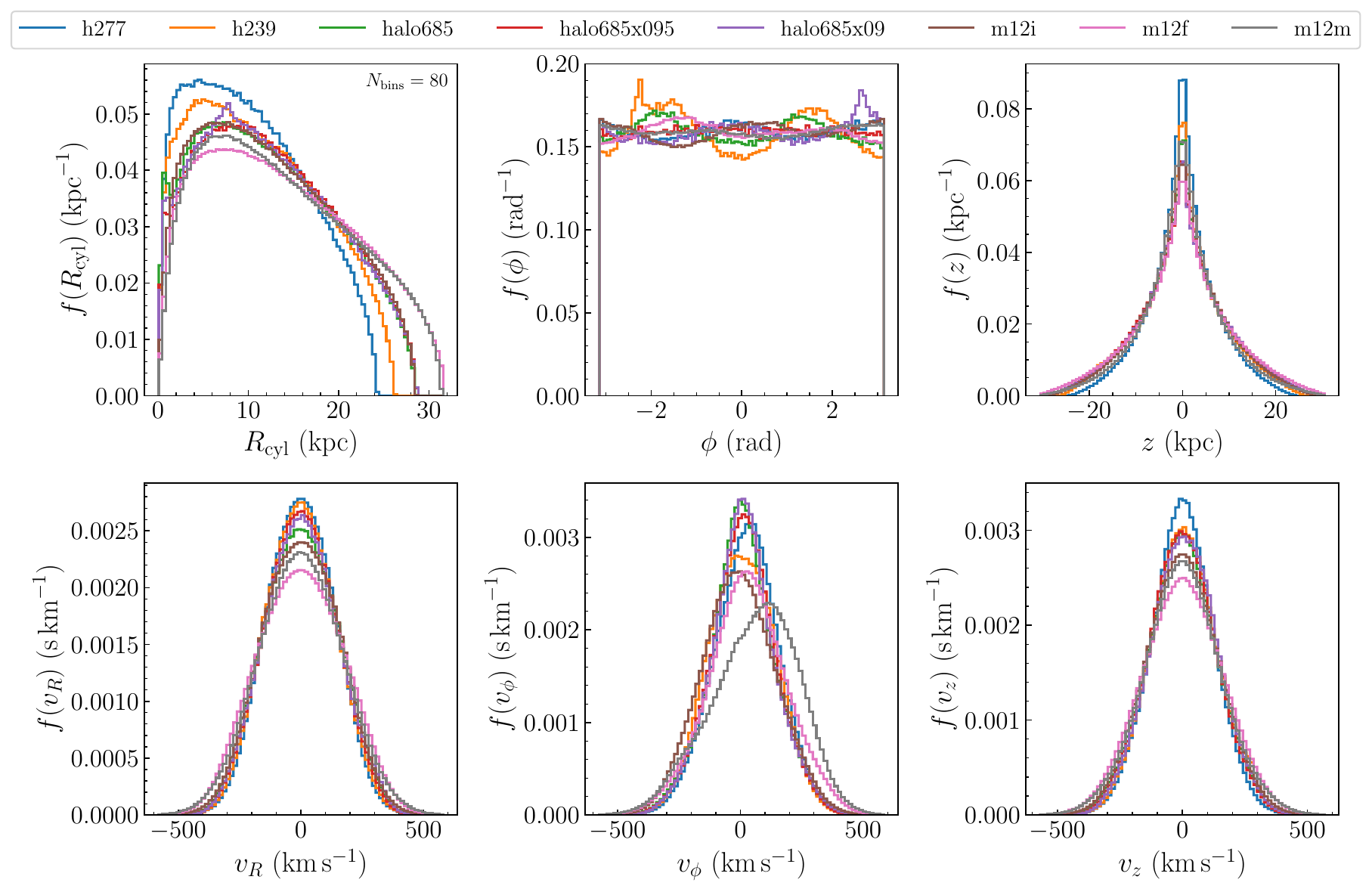}\label{fig:hist1d raw DM cylind}
    \vspace{-6mm}
    \caption{DM phase-space densities averaged within 28~kpc from the galactic centre in Cartesian and Cylindrical coordinates. Notice that all simulations have broadly similar distributions, apart from {\tt halo685x09} and {\tt m12m}. {\tt halo685x09} has a large DM subhalo $\sim 10$\,kpc from the Galactic centre in the process of merging (see the peak in the $f(R_{\rm cyl})$ panel). {\tt m12m} has a prominent accreted `dark disc,' showing significant rotation ($v_\phi \sim 150$\,km/s; see middle bottom panel). In the analysis which follows, we use DBSCAN to remove the subhalo from {\tt halo685x09}.}
    \label{fig:hist1d raw DM}
\end{figure}

In Figure~\ref{fig:hist1d raw DM}, we show the phase-space density of the dark matter in each galaxy within 28~kpc of the centre along each of the six components of position and velocity, in both Cartesian and cylindrical coordinates. We note two outliers: \texttt{halo685x09} and \texttt{m12m}.
\begin{itemize}
    \item\texttt{halo685x09} has a large DM subhalo located near $R_{\rm cyl}\sim10$~kpc and $\phi\sim3$ (in galactocentric cylindrical coordinates in Figure~\ref{fig:hist1d raw DM}). This subhalo is locally denser and faster than all other DM particles across the simulations, indicating that it is a genuine bound structure rather than a random fluctuation, and as such lies beyond the scope of our universal model. In this work we exclude this peaked DM and stellar substructure using a simple DBSCAN~\citep{1996kddm.conf..226E} clustering over 50 samples with a Euclidean metric in kinematic space, after which the resulting distributions are smooth and axisymmetric. The distribution of non-phase-mixed DM in Milky Way substructure has been studied in \citep{Necib:2018iwb,OHare:2019qxc}.

    \item \texttt{m12m} contains a component of dark matter in an oblate, spinning, ``dark disk''~\citep{Read:2008fh, Read_2009,Purcell_2009,Debattista_2019}. This feature can be seen in near $v_\phi\sim150$~km/s in Figure~\ref{fig:hist1d raw DM}, while  the rest of the simulated halos have $v_\phi\sim20$~km/s. 
    As we will discuss, the scaling strategy we develop will have difficulty generalizing to \texttt{m12m} due to the significant subcomponent of rotating dark matter.  
    However, inspecting the VPDF of the stars, we find that this dark disk is associated with a fast-spinning stellar disk. 
    This suggests that the presence or absence of a dark disk in our own Milky Way could be revealed through the motion of stars measured by {\em Gaia}~\citep{ 2016A&A...595A...1G,2021A&A...649A...1G}. We will discuss this in more detail in Section~\ref{sec:BDT}.

\end{itemize}

\subsection{Data Preprocessing and ROI Selection}

The goal of this work is to understand the phase-space distribution of the smooth DM component of Milky Way-size galaxies. Though the simulated galaxies are all ``Milky Way''-like, each differs slightly in mass. This global property sets an overall scale for the phase-space distributions. If uncorrected, these different scales would make comparisons between simulations (or with the Milky Way itself) impossible. 
Therefore, after the substructure exclusion, 
we follow the methodology inspired by \citet{Folsom:2025lly} and rescale the physical dimensions of each simulated halo to a common Milky Way baseline by resizing the virial radius of each halo, $R_{\mathrm{vir}}^{h_i}$, to the canonical Milky Way value, $R_{\mathrm{vir}}^\mathrm{MW} \equiv 200$~kpc. The resulting transformation of each phase-space coordinate $\mathbf{w} = (\vec{x},\vec{v})$ for each simulated halo $h_i$ is defined as
\begin{equation}\label{eq:virial rescaling}
    \vec{x} \rightarrow \vec{x} \cdot \frac{R_{\mathrm{vir}}^\mathrm{MW}}{R_{\mathrm{vir}}^{h_i}}, \qquad \vec{v} \rightarrow \vec{v} \cdot \sqrt{\frac{R_{\mathrm{vir}}^{h_i}}{R_{\mathrm{vir}}^\mathrm{MW}}}.
\end{equation}
We refer to this virial rescaling as \textbf{baseline} scaling, and use it to define the Region of Interest (ROI), subsequent data selection, and operations in physical units in this work. Table~\ref{tab:cuts_28kpc} overviews this and other data preprocessing we apply throughout the paper.

For diagnostics that require comparing radial or velocity profiles across halos on a common axis, we additionally introduce a per-halo \emph{uniform resizing} of the phase-space magnitudes within the ROI. For each halo $h_i$, let $r_\mathrm{max}^{h_i}$ and $v_\mathrm{max}^{h_i}$ denote the maximum galactocentric radius and speed of its DM particles after the virial rescaling of Eq.~\eqref{eq:virial rescaling}. We define the rescaled coordinates
\begin{equation}\label{eq:uniform rescaling}
    r' \equiv |\vec{x}| \cdot \frac{28~\mathrm{kpc}}{r_\mathrm{max}^{h_i}},
    \qquad
    v' \equiv |\vec{v}| \cdot \frac{544~\mathrm{km/s}}{v_\mathrm{max}^{h_i}},
\end{equation}
so that every halo's particles span $r' \in [0, 28]~\mathrm{kpc}$ and $v' \in [0, 544]~\mathrm{km/s}$ by construction. The reference values $28~\mathrm{kpc}$ and $544~\mathrm{km/s}$ correspond to the maxima of the baseline data for stars and DM within the galactic disks aggregated across the simulation suite. Unlike the virial rescaling of Eq.~\eqref{eq:virial rescaling}, which sets a common global potential depth, this uniform rescaling places all halos on a shared 
dynamic range and is used solely as a presentation device for the cross-halo profiles in Section~\ref{sec:BDT}. 

\begin{table}[tp]
\centering
\setlength{\tabcolsep}{8pt}
\begin{tabular}{lll}
\hline\hline
Step & Operation & Details \\
\hline
1 & Virial rescaling
  & $\mathbf{x} \to \mathbf{x}/(R_\mathrm{vir}/200)$,\
    $\mathbf{v} \to \mathbf{v}\,\sqrt{R_\mathrm{vir}/200}$ \\
2 & ROI selection & $(R_\mathrm{cyl} - R_\odot)^2 + z^2 \leq r_0^2$,\
    $R_\odot = 8.1$, $r_0 = 4.0~\mathrm{kpc}$ \\
3 & Subhalo exclusion
  & \texttt{halo685x09} only; HDBSCAN clustering \\
\hline
4a & {Baseline}
   & Virial rescale only\\
4b & {DM std}
   &  Standardize all 6D DM population \\
4c & Virial-stellar std
   & Virial radius rescale, scale velocity from ex-situ star reference \\
4d & Stellar std
   & Scale radius and velocity from ex-situ star reference \\
\hline\hline
\end{tabular}
\caption{Data processing pipeline for classification task in Section~\ref{sec:BDT} and generation task in Section~\ref{sec:flows}.
Steps~1--3 are shared by all scaling methods; Step~4 varies by standardization mode introduced in Eq.~\eqref{eq:DM std}: the baseline, moment-based standardization, virial-stellar moment-based standardization and stellar moment-based standardizations. }
\label{tab:cuts_28kpc}
\end{table}

To examine the universality of the DM phase-space in the volume of the simulated galaxies which correspond to observable regions of the Milky Way, we define our ROI so that it contains stars in the simulations which are comparable to the stars in the Milky Way close enough to the Sun that the {\em Gaia} Space Telescope can accurately measure their full position and velocity phase-space locations \citep{2016A&A...595A...1G}.  For this reason, our ROI is a toroidal volume in the plane of the galactic disk with a major radius of $8.1$~kpc (the distance of the Sun from the Galactic Center \citep{2019}) and a minor radius of $4$~kpc. These selections occur after the virial rescaling (Step 1 of Table~\ref{tab:cuts_28kpc}) to make all galaxies the size of the Milky Way. While the actual volume in which {\em Gaia} has accurate parallax, proper motion, and radial motion measurements for a sizeable uniformly-sampled population is a sphere of radius $\sim 4$~kpc centred on the Solar location rather than a torus around the Solar radius, the limited number of particles in the simulations requires us to aggregates multiple Solar-neighbourhoods at varying azimuthal angles in order to obtain reasonable particle counts.  
Table~\ref{tab:Ndata} describes the statistics of the ROI. 

\begin{figure}[tp]
    \centering
    \includegraphics[width=.9\columnwidth]{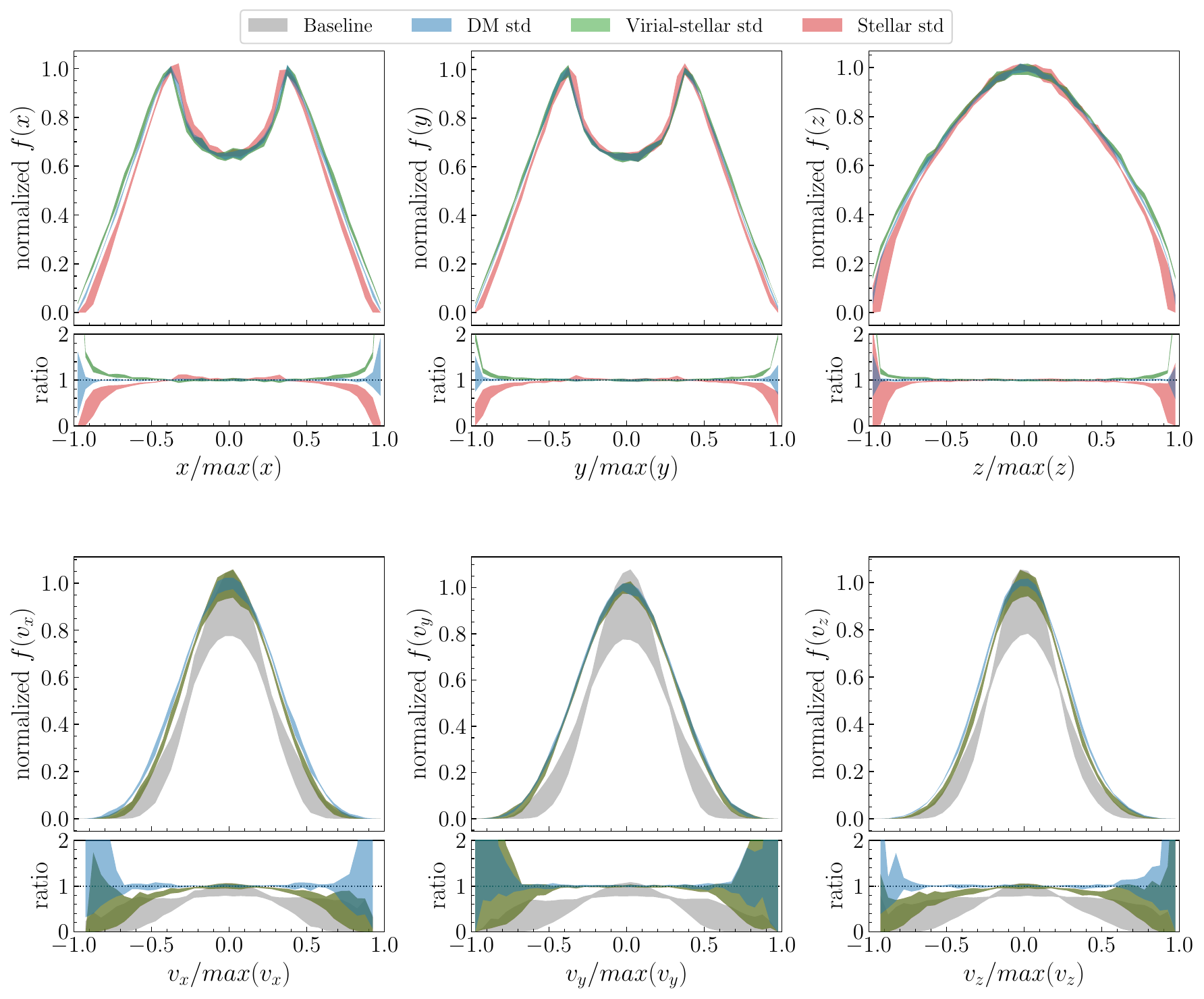}
    \caption{1D marginal histograms of DM position and velocity for the baseline, DM, virial-stellar, and stellar moment-based standardization in the ROI. The shading indicates the variation of each profile averaging over all halos. Note that the histogram of virial-stellar standardization method is overlapped with the baseline and stellar standardizations in position and velocity, respectively.
    }
    \label{fig:hist1D|DM std}
\end{figure}

In Figure~\ref{fig:hist1D|DM std}, the one-dimensional phase-space distributions in Cartesian coordinates for the ROI are shown in the grey bands, after applying the virial rescaling of Eq.~\eqref{eq:virial rescaling} to the raw data. After this rescaling, the position-space distributions differ only slightly across the simulation suite, but significant discrepancies persist in the velocity distributions, visible by eye. Rescaling by the virial radius thus normalizes the global potential depth but does not remove all structural differences between halos. These residual differences mean the halos remain out-of-distribution (OoD) relative to one another, so a model trained on one halo set generalizes poorly to the other. In the following section, we identify a more extensive statistical scaling that maps the halos onto a common, universal distribution so that a model trained on any subset generalizes to the rest.

\begin{table}[t]
\centering
\renewcommand{\arraystretch}{1.3}
\setlength{\tabcolsep}{5pt}
\resizebox{\textwidth}{!}{%
\begin{tabular}{l | c c c c c c c c }
\hline\hline
\textbf{Halo ID} & \texttt{h277} & \texttt{h239} & \texttt{halo685} & \texttt{halo685x09} & \texttt{halo685x095} & \texttt{m12i} & \texttt{m12f} & \texttt{m12m} \\
\hline
 \textbf{Stars} & 16.36k & 8.80k  & 1.28k & 1.17k & 1.18k & 20.64k & 12.43k & 22.11k \\
 \textbf{DM}    & 0.15M  & 0.13M  & 0.11M & 0.11M & 0.11M & 0.67M  & 0.98M  & 1.23M  \\
\hline\hline
\end{tabular}%
}
\caption{Number of star particles and DM particles in the simulated halos in the toroidal ROI. For the following multi-halo analysis, the particle number is balanced across halos.}
\label{tab:Ndata}
\end{table}

\section{A Universal Scaling Strategy}~\label{sec:classifier}

To rescale the disparate dark matter distributions to a single ``universal'' phase-space density that can be applied to all Milky Way-like spiral galaxies, we consider the hierarchy of statistical moments of the density distributions, taking the mean $\mu_{\mathrm{DM}}$ and standard deviation $\sigma_{\rm DM}$ of each coordinate in the six-dimensional position/velocity dataset in Cartesian coordinates after applying the virial rescaling Eq.~\eqref{eq:virial rescaling}. Our {\em ansatz} is that the shared kinematic structure of the different galaxies can be revealed by standardizing the first and second moments of their six-dimensional phase-space. This is consistent with recent findings that the dominant diffuse DM component from FIRE galaxies is well-described by a generalized Gaussian velocity distribution~\citep{Zhang:2026qnl}.
We perform this standardization by shifting each phase-space coordinate $\mathbf{w}$ of the dark matter distribution by that variable's first moment and scaling it by its second central moment:
\begin{equation}\label{eq:DM std}
    \tilde{w} = \frac{w - \mu_{\mathrm{DM}}}{\sigma_{\mathrm{DM}}}.
\end{equation}
This \textbf{(DM) moment-based standardization} removes trivial offsets in location and scale: any remaining residuals are discrepancies in the intrinsic ``shape'' (higher-order moments) of the phase-space distribution. Consequently, significant deviations in standardized profiles signal fundamental non-universality, identifying structural features distinct to specific hydrodynamic feedback implementations. Note that each coordinate $w$ becomes a dimensionless quantity $\tilde{w}$. 

Figure~\ref{fig:hist1D|DM std} demonstrates the efficacy of this approach in the ROI using one dimensional profiles of the phase-space. In contrast to the baseline method 
which only has the virial rescaling, within the ROI our moment-based standardized marginal distributions exhibit strong similarity across simulations.

In observational contexts, the true DM distribution is unobservable, rendering the direct calculation of its scaling parameters -- $\mu_\mathrm{DM}$ and $\sigma_\mathrm{DM}$ in Eq.~\eqref{eq:DM std} -- impossible. However, as noted in \citep{Necib:2018igl}, dark matter and stars accrete jointly into the host galaxy and relax over time. Because these stars formed ex-situ and were accreted alongside DM during the galaxy's assembly, they settle into the halo rather than the rotation-dominated disk. Consequently, their phase-space behaviour is expected to closely mimic the dispersion-dominated nature of the DM halo, rendering them an ideal dynamical proxy for normalizing the dark matter scale and allowing an approximation of the moments used in Eq.~\eqref{eq:DM std}.

\begin{figure}[t!]
    \centering
    \includegraphics[height=.35\textheight]{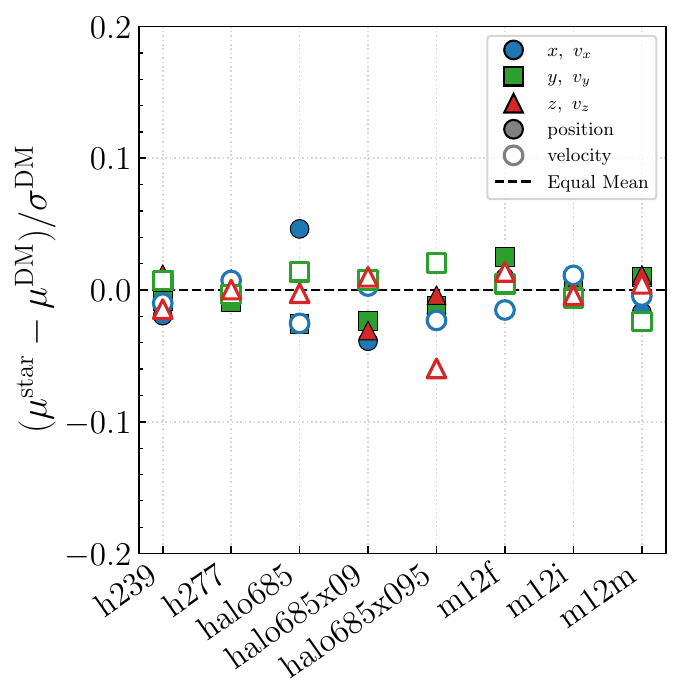}
    \includegraphics[height=.35\textheight]{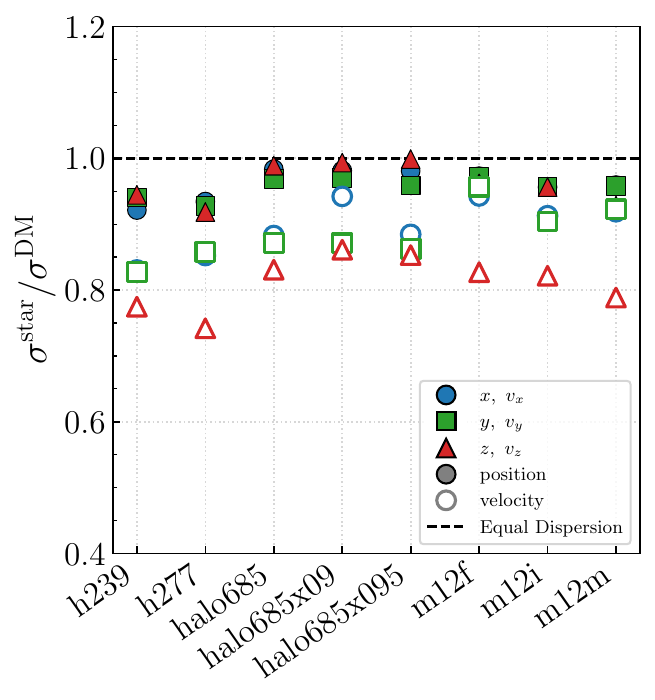}
    \caption{Ratio of stellar to DM relative mean $(\mu^\mathrm{star} - \mu^\mathrm{DM}) / \sigma^\mathrm{DM}$ and dispersion ($\sigma^\mathrm{star} / \sigma^{\mathrm{DM}}$) of the 6D variables in Cartesian coordinates across the simulation suite. The selected metallicity cut consistently recovers $\sim 90\%$ of the intrinsic DM dispersion, and the difference in mean values between the two distributions is negligible relative to the dispersion.}
    \label{fig:stats|DM-S}
\end{figure}

This accreted stellar population is predominantly metal-poor, and so we isolate these tracers by enforcing a metallicity cut. Empirically, we find in simulation that a threshold of $\log_{10}[\mathrm{Fe/H}] < -2$ optimizes the trade-off between statistical power and kinematic fidelity.\footnote{Note that the number of star particles in a simulation is significantly smaller than the number of stars in a galaxy, so in real data a different metallicity cut may provide a more optimal trade-off.} As shown in Figure~\ref{fig:stats|DM-S}, this selection criterion yields a small difference between the means of the two distributions in both position and velocity phase-space. The dispersion ratio of stars to DM yield a 5\% variation in a position axis, and in velocity axis it is about 8\%. The small relative mean difference and consistent dispersion ratio allow us to use the metal-poor stars as a robust, calibrated proxy for the unobservable DM halo.

Building on this proxy, we propose two methods to standardize the coordinates that differ in how much stellar information they rely on. The first, the {\bf virial-stellar moment-based standardization}, is a hybrid construction. The positional variables are rescaled by the the virial radius as in Eq.~\eqref{eq:virial rescaling}. The velocity components follow the standardization of Eq.~\eqref{eq:DM std}, with the dark-matter moments replaced by their stellar counterparts, $\mu_\mathrm{DM}\rightarrow\mu_\mathrm{star}$ and $\sigma_\mathrm{DM}\rightarrow\sigma_\mathrm{star}$, where $\mu_\mathrm{star}$ and $\sigma_\mathrm{star}$ are the mean and standard deviation of the low-metallicity stellar population in Cartesian coordinates. The resulting transformed velocity coordinates in Figure~\ref{fig:hist1D|DM std}. This mode inherits the strong recovery of the Milky-Way size from the baseline mode, but relies on a precise estimate of the inner-galaxy mass, which carries an error of around 30\% for the Milky Way~\citep{Posti_2019,Watkins_2019,2022ApJ...925....1S,2024ApJ...972...70R}.

The second method, {\bf stellar moment-based standardization}, extends this standardization to all six phase-space coordinates, applying it to position as well as velocity. This avoids the reliance on the enclosed mass or virial radius, trading it for two position-based error sources: the {\em Gaia} positional measurements, which contribute roughly 10\% depending on the distance along the line of sight~\citep{2021A&A...649A...1G}, and the halo-to-halo variance in the star-to-DM position dispersion shown in Figure~\ref{fig:stats|DM-S}, which contributes a further $\sim$10\%. The resulting distribution is shown in Figure~\ref{fig:hist1D|DM std}, alongside the other standardizations.

\section{Validating Universality via Two-Sample Classifier Tests}\label{sec:BDT}

While one-dimensional marginals allow for straightforward visual inspection, they cannot reveal the full six dimensional phase-space structure.  To rigorously study the degree of similarity of the phase-space distributions across simulations, we employ a two-sample classifier test. A binary classifier on the high-dimensional phase-space for two sets of particles (``Class-0'' and ``Class-1''). By the Neyman-Pearson Lemma \cite{10.1098/rsta.1933.0009}, the optimal classifier is monotonic with the ratio between the two phase space densities,
\begin{equation}
R_{\rm optimal}(\tilde{\mathbf{w}}) = {p_1(\tilde{\mathbf{w}})\over p_0(\tilde{\mathbf{w}})}
\end{equation}
A well-trained near-optimal classifier should approximate this $R_{\rm clf}(\tilde{\mathbf{w}})\approx R_{\rm optimal}(\tilde{\mathbf{w}})$. Therefore, if $R_{\rm clf}(\tilde{\mathbf{w}})$ is close to 1 everywhere, the two datasets have approximately the same phase space density. The two-sample classifier test is a stringent test of the similarity of two datasets phase space densities that captures all of the correlations between the six phase space coordinates.

The particular two-sample test we employ is a One-vs-Rest Gradient Boosted Decision Tree (BDT) classifier \citep[{LightGBM};][]{lightGBM} to distinguish the dark matter particles in a target galaxy (Class-1) from the aggregate ensemble of dark matter particles from all the other galaxies (Class-0). We iteratively cycle through our simulated galaxies, treating each in turn as the target Class-1 to be compared against the remaining ensemble using the standardized phase-space location $\mathbf{\tilde{w}}$ as defined in Eq.~\eqref{eq:DM std} (after either the DM, stellar, or virial-stellar standardization modes).

We choose to assess the global performance of the classifier via the AUC metric of the Receiver Operating Characteristic (ROC) curve. The AUC is the integral of the true positive rate (TPR) as a function of the false positive rate (FPR); an AUC of 1 indicates perfect separation of the two classes, while an AUC of 0.5 means that classifier is incapable of distinguishing the datasets.

\begin{figure}[tp]
    \centering
    \includegraphics[width=1.\linewidth]{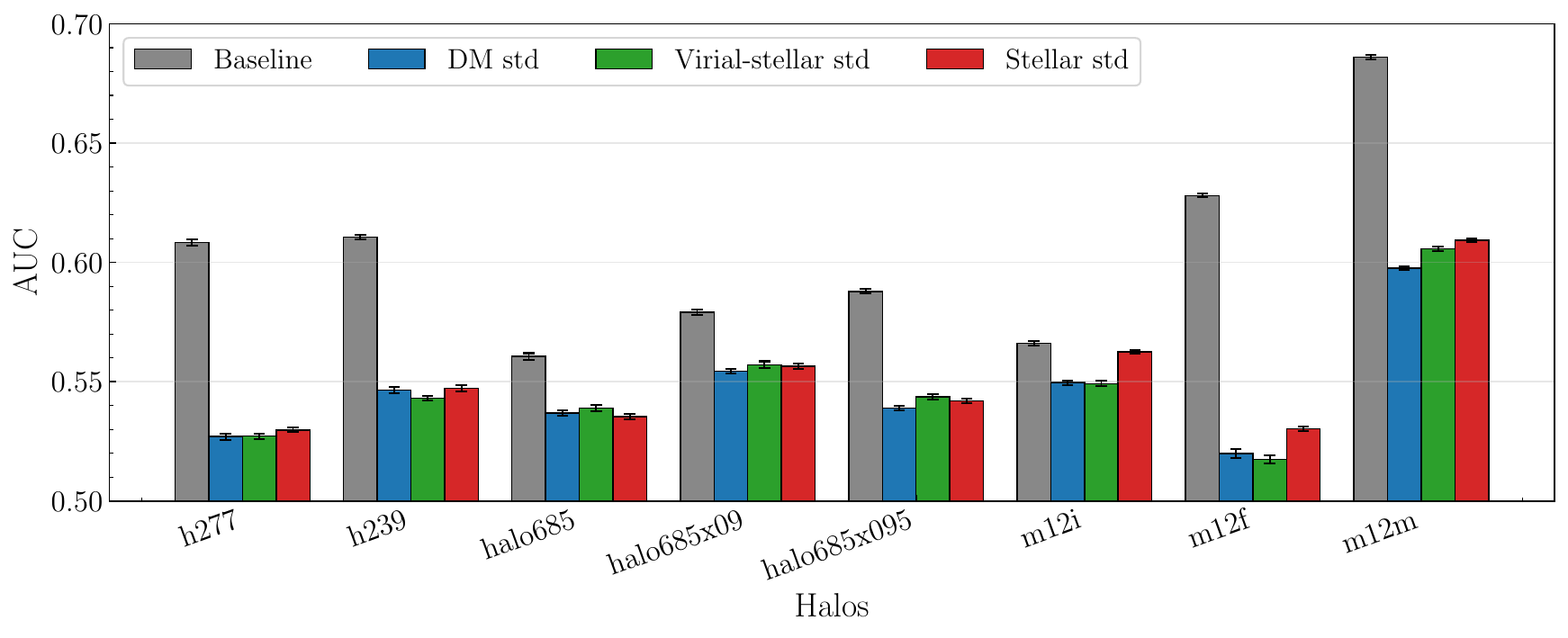}
    \caption{Overall bootstrap AUC for each halo under the four standardization modes: Baseline (no standardization, gray), DM-based (blue), virial-stellar (green), and stellar standardization (red). The classifier separates one halo from the aggregated halo built from the rest: a lower AUC means the standardization is more universal. Error bars are the bootstrap uncertainties.}
    \label{fig:compare scaling AUC overall-annulus}
\end{figure}

Figure~\ref{fig:compare scaling AUC overall-annulus} shows the AUC score of each halo's particles in the ROI for the four standardization modes summarized in Table~\ref{tab:cuts_28kpc}. The moment-based standardizations significantly lower the AUC by enhancing the cross-galaxy similarity relative to the baseline mode. The DM standardization (blue) and the viral-stellar standardized dataset (green) track each other closely, typically within ${\cal O}(0.1\%)$, confirming that stellar moments are an effective proxy for the DM velocity distribution. The full 6D stellar standardized dataset (red) in most cases closely follows the other two standardizations; though it is noticeably weaker for the \texttt{m12i} and \texttt{m12f} halos. 

\begin{figure}[tp]
    \centering
    \includegraphics[width=0.65\linewidth]{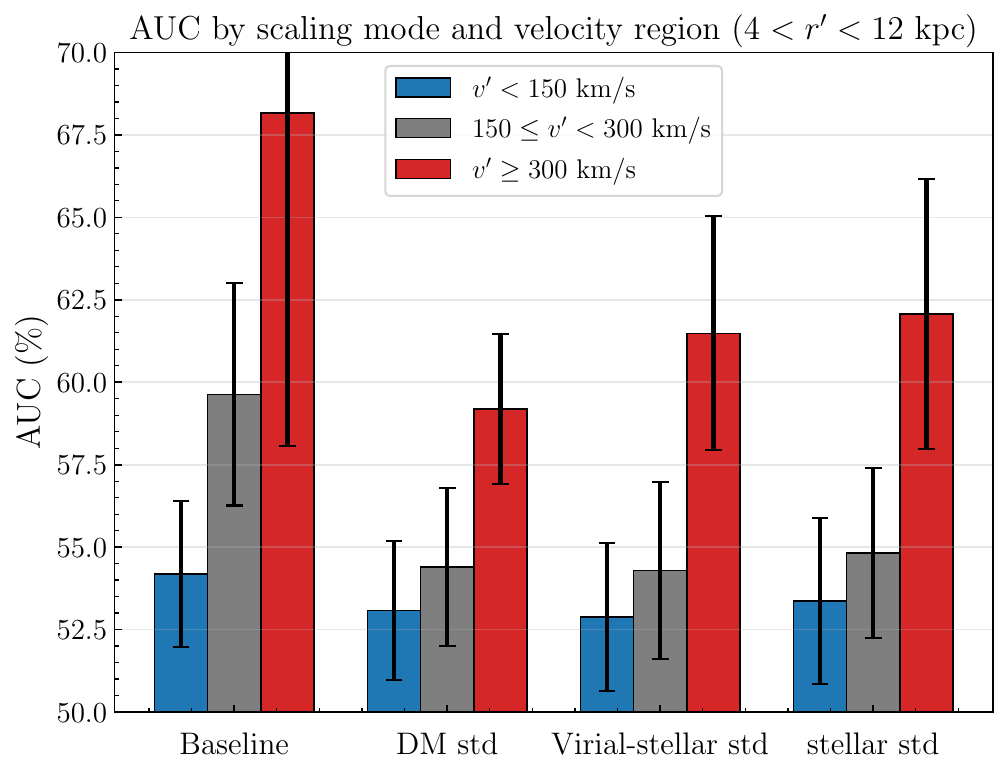}
    \caption{Mean AUC over halos for each standardization mode (Baseline, DM-based, virial-stellar, and stellar), broken down by velocity bins: 
    $v^\prime<150$ km/s, $150\le v^\prime<300$ km/s, and $v^\prime\ge300$ km/s, within $4<r^\prime<12$ kpc. A lower AUC indicates a more effective scaling.  (An AUC of 1 indicates perfect separation of the two classes, which in this case is a failure mode, while an AUC of 0.5 means that classifier is incapable of distinguishing the datasets, which in this case is ideal behaviour.) Error bars reflect variation between halos.}
    \label{fig:regional AUC comparison}
\end{figure}

To more closely interrogate the success of this standardization, in Figure~\ref{fig:regional AUC comparison} we consider the AUC of particles within the ROI averaged over all halos, binned in resized velocity $v'$ following Eq~\eqref{eq:uniform rescaling}. We see that low- and intermediate velocities ($v' < 150$~km/s and $150\leq v' \leq 300$~km/s, respectively) have low AUCs (less than 55\% averaged over all halos) using all standardization modes, while the AUCs of the high velocity tail of the distribution is consistently higher for all standardizations.

\begin{figure}[t!]
    \centering
    \includegraphics[width=0.9\columnwidth]{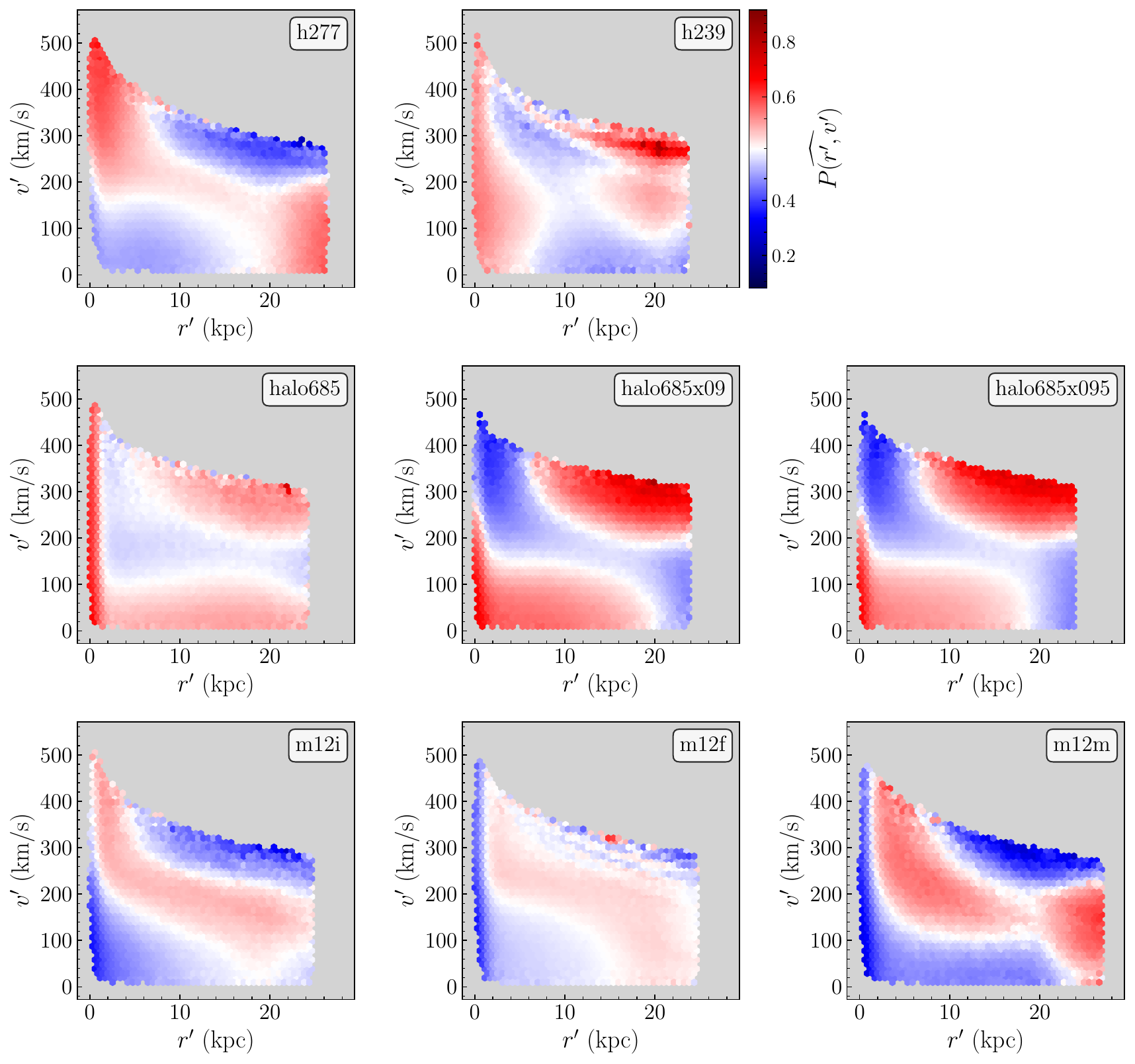}
    \caption{DM moment-based standardized phase-space heatmaps of $\widehat{P\left(r^\prime, v^\prime\right)}$ for each halo. Dark red and blue indicate the regions where the greatest deviations from universality occur, while white indicates ideal behaviour. The coordinates are under uniform rescaling defined in Eq.~\eqref{eq:uniform rescaling}.}
    \label{fig:BDT2D-DM_std_cartesian}
\end{figure}

We can further resolve the phase-space locations where the universality found in the DM moment-based standardization breaks down by applying the classifier to particles outside of the toroidal ROI.
The BDT returns the classifier probability $P_c(\mathbf{\tilde{w}})$, which is the estimate that a particle at phase-space location $\mathbf{\tilde{w}}$ belongs to Class-1 rather than Class-0. 
We summarize the per-particle probability along a chosen set of phase-space coordinates $\xi$ (e.g., $\xi = r^\prime$ or/and $v^\prime$), and report the binned median (averaging over the other phase-space coordinates within each bin of $\xi$)
\begin{equation}
    \widehat{P(\xi)} \equiv \mathrm{median}\!\left\{P_c(\xi_i) : \xi_i \in \mathrm{bin}(\xi)\right\}.
    \label{eq:cond-entropy}
\end{equation}
In Figure~\ref{fig:BDT2D-DM_std_cartesian} we map $\widehat{P(r^\prime,v^\prime)}$ with the DM moment-based standardization for each halo across all radii of the galaxies' disks, $r' < 28$~kpc .\footnote{$r^\prime$ and $v^\prime$ are defined in Eq.~\eqref{eq:uniform rescaling}.} Regions where $\widehat{P(r^\prime,v^\prime)} \approx 0.5$ (white, pale blue, pale red) mark regions of maximum indistinguishability. These regions cover the virialized bulk: $r^\prime \in (4, 20)~\mathrm{kpc}$ at intermediate speeds $v^\prime \in (100, 300)~\mathrm{km/s}$, in agreement with the AUC values seen in the toroidal ROI. Conversely, the saturated red and blue regions highlight non-universal substructures that moment-based scaling cannot remove. For example, \texttt{h277} contains a high-velocity $\widehat{P}>0.5$ plume at $r^\prime > 20~\mathrm{kpc}$, identifying unrelaxed merger debris in its outskirts. \texttt{halo685} has a dense high-probability core at $r^\prime < 2~\mathrm{kpc}$, indicating a cuspier inner density profile than the ensemble average. \texttt{m12i} exhibits a dipole in the inner halo --- a excess of high velocity particles against a background-like low-velocity population (blue) --- reflecting its specific mass concentration. 

While collectively these maps confirm that the standardization establishes a universal baseline within the virialized bulk where the toroidal ROI (and the spherical region where {\em Gaia} measurements exist in the Milky Way) is located, there is an notable and interesting exception. The simulation \texttt{m12m} is an outlier across all of our standardization modes (though all are improvements over the baseline). 
This exceptionalism appears to be the result of a co-rotating dark disk, which is visible in Figure~\ref{fig:hist1d raw DM} as the offset peak in $f(\tilde{v}_\phi)$. This bulk rotation is a halo-specific feature largely absent from the other halos, and so serves as an easily-identified feature by the classifier, reflected in the higher AUC for \texttt{m12m} when compared to the other halos even after standardization in Figure~\ref{fig:compare scaling AUC overall-annulus}. 

\section{A Normalizing Flow Model of the Universal DM Distribution}~\label{sec:flows}

Thus far, we have demonstrated that a simple scaling strategy allows us to standardize the DM distributions of galaxies, as measured by a classifier trained on the full dimensionality of the phase-space. We now go further to create a numeric model of this universal distribution, using deep learning algorithms. In addition to providing an additional window onto the universal scaling strategy, such a generative model will allow numeric calculation of the local dark matter phase-space density when applied to real {\em Gaia} data in future work.
We utilize Masked Autoregressive Flows (MAFs)~\citep{papamakarios2018maskedautoregressiveflowdensity}, which are unsupervised algorithms that learn the probability density of a target dataset. These algorithms are also generative, allowing us to create synthetic DM particles sampled from the learned distribution.

MAFs parametrize a bijective mapping $f: \mathbf{u} \rightarrow \mathbf{w}$ between a simple latent base distribution $\pi(\mathbf{u})$ (typically a standard isotropic multidimensional Gaussian) and the complex target DM phase-space $p(\mathbf{w})$. The probability density of the target data is evaluated via the change of variables formula:
\begin{equation}
    \log p(\mathbf{w}_\mathrm{DM}) = \log \pi(\mathbf{u}) + \log \left| \det \frac{\partial f^{-1}}{\partial \mathbf w} \right|,
\end{equation}
where the autoregressive structure of the MAF ensures that the Jacobian matrix is triangular, rendering the determinant computation efficient. This architecture offers a dual advantage: it allows for exact likelihood estimation (i.e., estimation of the phase-space density) and high-fidelity data sampling.

We train our MAF on the scaled positions and velocities $\tilde{\mathbf{w}}$ of the DM particles in the stacked catalogue of all simulated halos. As a result, the network effectively learns the ensemble-averaged phase-space distribution of the entire population. While the density estimator provides statistical likelihoods, we validate the reconstruction of the ``universal'' halo profile via the models' generative capability trained by four different scalings list in Step~4 of Table~\ref{tab:cuts_28kpc}: baseline, DM standardized, virial-stellar standardized, and stellar standardized halos. Comparing samples generated by this ensemble-trained model against the ground truth of a specific halo provides another measure of the success of our scaling Eq.~\eqref{eq:DM std} in creating a universal profile across simulated galaxies that boost the the domain adaption. A high degree of similarity implies that the scaling has successfully minimized inter-galactic structural variance.

To improve the MAF training, we adopt a refined preprocessing pipeline based on \citep{Buckley:2022tjy}, applying centralization, a logit transformation, and standardization. First, to handle the finite boundaries of the ROI, we map the ensemble of scaled galaxies into a unit ball via:
\begin{equation}
    \vec{x} \rightarrow \frac{\vec{x} - \vec{x}_\odot}{r_\mathrm{max}\cdot (1+\epsilon)},
\end{equation}
where position inputs $x$ drawn from the relevant dataset ({i.e.},  baseline, DM, virial-stellar, or stellar standardized) depending on the standardization mode,  $\vec{x}_\odot$ represents the geometric center, and $\epsilon$ is a small margin parameter creating a buffer between the dataset and the boundary radius $r_\mathrm{max}$. To map this bounded domain to the full real space $\mathbb{R}^3$ required by the MAF, we apply a logit transformation:
\begin{equation}
    \vec{x} \rightarrow \frac{\vec{x}}{|\vec{x}|} \tanh^{-1}|\vec{x}|.
\end{equation}
This projects the radial component from the unit sphere onto the real line, thereby mitigating boundary artifacts associated with the hard truncation of the ROI. Finally, the velocities and transformed positions are standardized to ensure numerical stability during training.

\begin{figure}[tp]

    \includegraphics[width=1.\columnwidth]{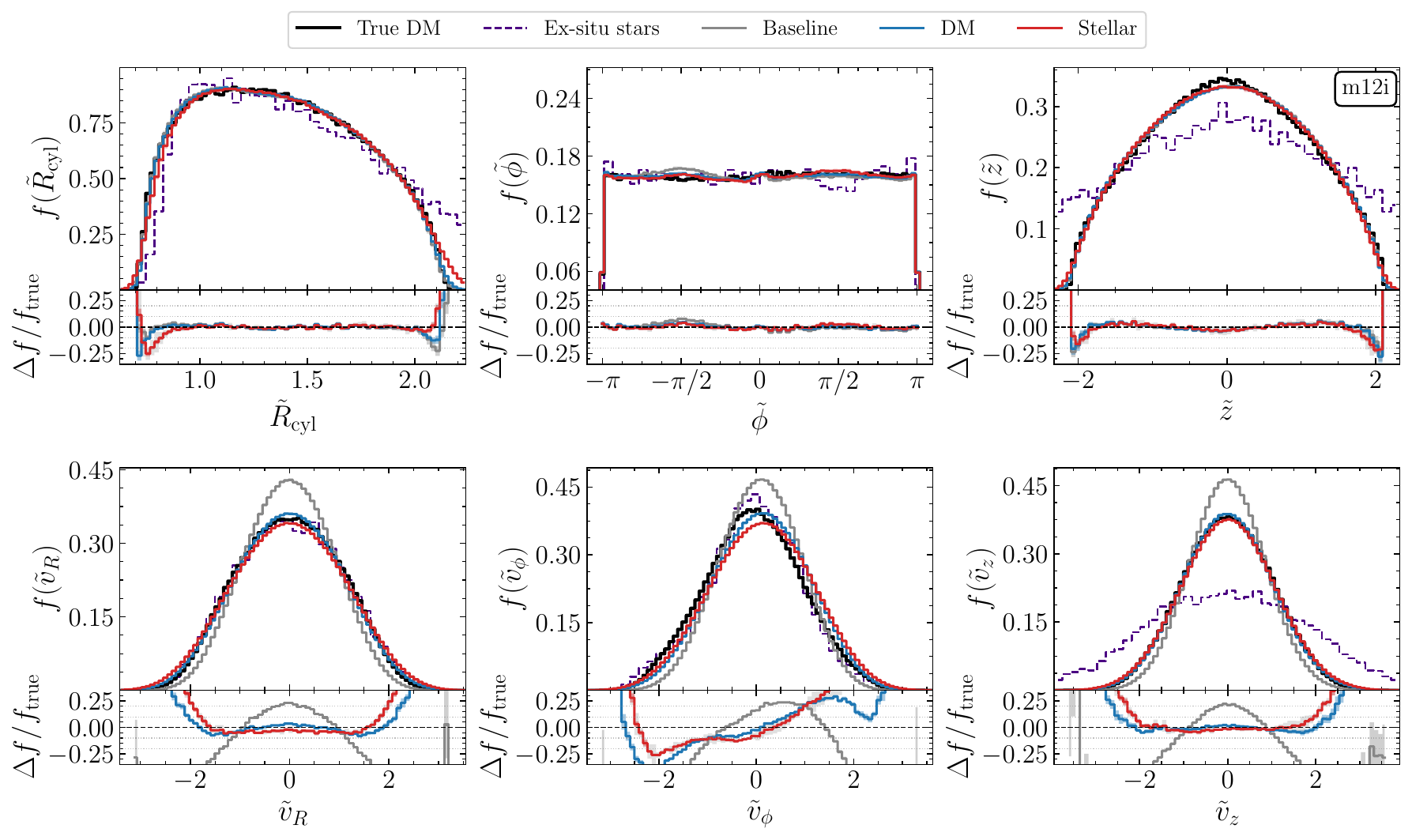}
    \caption{Reconstructed 1D phase-space histograms of the \texttt{m12i} halo in the ROI from a generative model trained on Training~Set~A in Figure~\ref{fig:AUC_results} in cylindrical coordinates. Top and bottom rows show position ($\tilde{R}_{\rm cyl},\tilde{\phi},\tilde{z}$) and velocity ($\tilde{v}_R,\tilde{v}_\phi,\tilde{v}_z$) distributions, respectively, with sub-panels displaying the fractional residual ($\Delta f/f_\mathrm{true}$) relative to the ground truth (black).
    The DM- (blue) and stellar standardized (red) modes track the truth closely, while the baseline (gray) distributions show clear residual biases. The distribution of ex-situ stars (purple dashed) is overlaid to visualize its rotational signature in $\tilde{v}_\phi$ and its overall shape relative to the DM.}
    \label{fig:NFhist1D|DM-S-cyl}
\end{figure}

We first qualitative evaluate the generative fidelity by comparing the flow-generated predictions directly against the ground truth for an example halo. Figure~\ref{fig:NFhist1D|DM-S-cyl} shows the reconstructed phase-space distributions of the \texttt{m12i} halo from a model trained on \texttt{h277}, \texttt{h239} (N-Body Shop), \texttt{halo685} and \texttt{halo685x095} (VINTERGATAN-GM), expressed in units of the standardized true DM distribution. The necessity of standardization is immediately apparent in the velocity sectors (bottom rows), where the unscaled generation (gray) fails to match the target dispersion, yielding residuals exceeding $25\%$. In contrast, the standardized models (blue and red) achieve remarkable fidelity, maintaining spatial matching precision within $5\%$ throughout the ROI and velocity reconstruction accuracy of $\sim 10\%$ within the distribution bulk in both coordinate systems.

Crucially, while the residuals for the scaled models largely remain within the intrinsic variance of the training data even in the core, they exceed this resolution limit in the high-velocity tails. This divergence not a generative failure: it confirms that while the thermalized bulk follows a universal scaling, the high-velocity outliers are dominated by stochastic, halo-specific merger histories that cannot (even in principle) be perfectly inferred from a generalized training ensemble. 

\begin{figure}[tp]
    \centering
    \includegraphics[width=1.\linewidth]{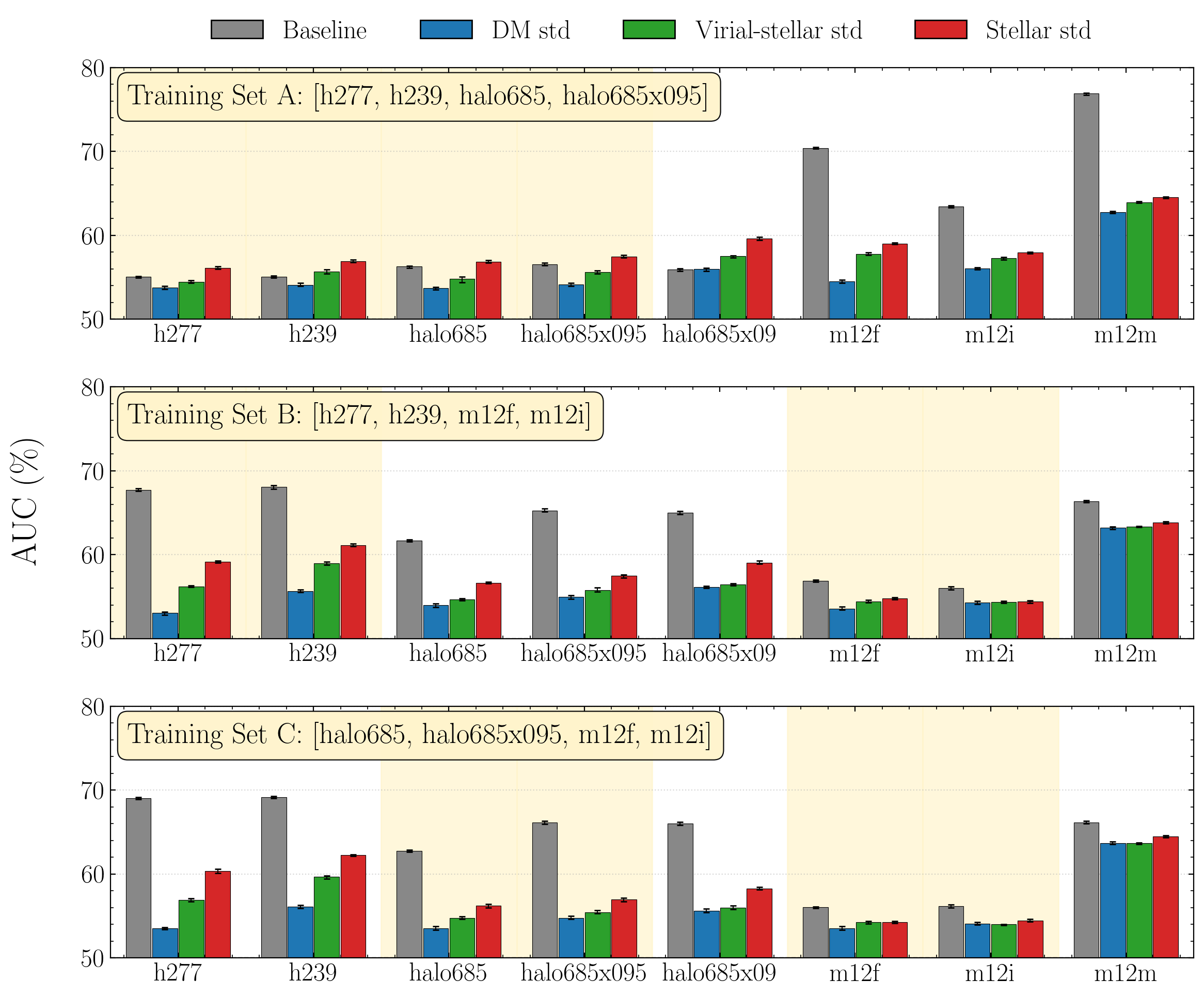}
    \caption{Charts of classification results between truth and generative halos from flow models trained with two simulation suites in all combinations. The yellow bands indicate the galaxies in the training data. 
    }
    \label{fig:AUC_results}
\end{figure}

To quantify the generative fidelity across the full 6D phase-space regardless of the choice of coordinates, we again employ a BDT classifier to discriminate between the generated samples and the true DM distribution. 
Figure~\ref{fig:AUC_results} presents the resulting AUC scores for all cross-validation combinations among the training suites, where the flow model is trained on scaled particles from a subset of the total available halos and tested on the remaining galaxies for examination of transfer learning. 
The three training sets are: 
\begin{enumerate}[label=(\Alph*)]
\item Nbody Shop (\texttt{h277} \& \texttt{h239}) + Vintergatan-GM (\texttt{halo685} \& \texttt{halo685x095}),
\item Nbody Shop (\texttt{h277} \& \texttt{h239}) + FIRE-2 (\texttt{m12i} \& \texttt{m12f}),
\item Vintergatan-GM (\texttt{halo685} \& \texttt{halo685x095}) + FIRE-2 (\texttt{m12i} \& \texttt{m12f}).
\end{enumerate}
The results consistently demonstrate that the DM-scaled data significantly outperform the unscaled baselines, yielding AUC scores close to 0.5, followed by the stellar scaled data with a small degradation in performance. 

Table~\ref{tab:flow_moments} shows the recovery of the mean and standard deviation of DM speed distribution at the Solar location given the stellar moment-based standardization results of Training~Set~A. On the training halos, the flow reproduces both speed moments with mean absolute errors of $\sim4$\% in $\mu_v$ and $\sim3$\% in $\sigma_v$. As expected, generalization to the test halos (\texttt{halo685x09}, \texttt{m12i}, and \texttt{m12f}) is weaker, with mean absolute errors rising to $\sim6$\% in $\mu_v$ and $\sim8$\% in $\sigma_v$. The performance on the two \texttt{FIRE-2} galaxies \texttt{m12i} and \texttt{m12f} is notably worse than for the held out Vintergatan-GM galaxy \texttt{halo685x09}, which likely reflects the cost of the greater domain shift, since Training~Set~A contains two other  Vintergatan-GM galaxies but no  \texttt{FIRE-2} galaxies. But overall, generalization performance at the $\sim 5-10\%$ level is quite good in absolute terms and is a promising sign for the future application to real data. 

\begin{table*}[t]
\centering
\renewcommand{\arraystretch}{1.2}
\setlength{\tabcolsep}{7pt}
\begin{tabular}{l c c  c c }
\hline\hline
 & \multicolumn{2}{c}{Truth (km/s)} & \multicolumn{2}{c}{Flow $-$ Truth (\%)} \\
Halo & $\mu_v$ & $\sigma_v$ & $\Delta\mu_v/\mu_v$ & $\Delta\sigma_v/\sigma_v$  \\
\hline
\texttt{h277}        & 221.6 & 81.7  & $-3.2$ & $+3.2$  \\
\texttt{h239}        & 222.8 & 87.9  & $-3.2$ & $-3.8$    \\
\texttt{halo685}     & 244.1 & 96.1  & $+3.3$ & $+2.7$    \\
\texttt{halo685x095} & 233.1 & 95.3  & $+4.6$ & $+0.4$   \\
\hline
\textbf{mean absolute error} &     &       & 3.6 & 2.5    \\
\hline
\texttt{halo685x09}  & 231.1 & 95.3  & $+6.2$ & $+1.0$    \\
\texttt{m12i}        & 274.7 & 100.5 & $+4.1$ & $+11.4$   \\
\texttt{m12f}        & 313.9 & 120.0 & $+8.6$ & $+11.4$   \\
\hline
\textbf{mean absolute error} &     &       & 6.3 & 7.9    \\
\hline
\texttt{m12m}         & 333.8 & 117.2 & $+3.5$ & $+16.3$  \\
\hline\hline
\end{tabular}
\caption{Recovery of the DM speed-distribution moments under stellar moment-based standardization at the Solar location. Columns give the ground-truth mean speed $\mu_v$ and dispersion $\sigma_v$, followed by the fractional errors of our reconstruction of Training~Set~A. The upper block contains the training halos and the lower block the test ones, each followed by its mean absolute error; the final row shows the outlier \texttt{m12m}. The flow reproduces the speed moments to a few percent for the train halos, while the dispersion error grows to $\sim10\%$ for \texttt{halo685x09}, \texttt{m12i}, \texttt{m12f}, and especially \texttt{m12m}, whose accreted co-rotating dark disc broadens and shifts the speed distribution in a halo-specific way.}
\label{tab:flow_moments}
\end{table*}

\begin{figure}[tp]
    \includegraphics[width=1.\textwidth]{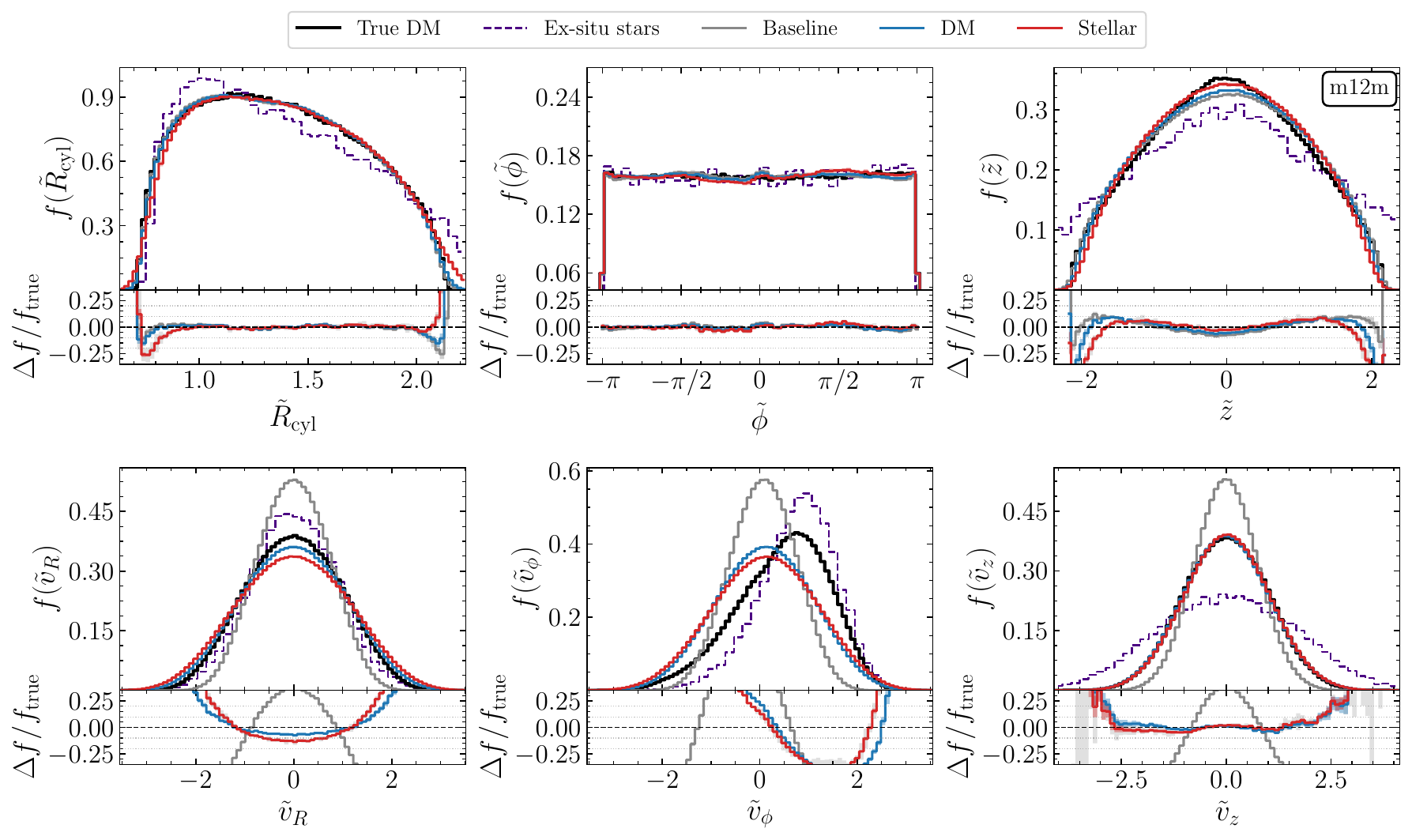}
    \caption{Reconstructed 1D phase-space histograms of the \texttt{m12m} halo in cylindrical coordinates. This halo hosts a co-rotating dark disk, evident as the offset peak in $f(\tilde{v}_\phi)$, whose rotation does not align with that of the ex-situ stars (purple dashed). None of the standardizations fully recover this rotational feature: the DM- (blue) and stellar standardized (red) modes underestimate the shift, while the baseline (gray) distributions remain clearly biased.}
    \label{fig:NFhist1D|DM-S-cyl-m12m}
\end{figure}

The last line of Table~\ref{tab:flow_moments} shows the performance of the universal model on the outlier galaxy \texttt{m12m}. Here the prediction of the DM velocity mean at the Solar location is actually not bad, but the prediction of the DM velocity standard deviation is considerably worse, rising to 16\% deviation. This failure can again be attributed to \texttt{m12m}'s dark disk rotation as shown in Figure~\ref{fig:NFhist1D|DM-S-cyl-m12m} and Table~\ref{tab:flow_moments}. In Figure~\ref{fig:NFhist1D|DM-S-cyl-m12m}, despite the DM and the ex-situ halo stars both rotating (though the stars are off-set from the DM), all generated DM halos remain centred at zero due to the lack of similar rotation present in the other galaxies in the training set.
The unscaled baseline suffers a significant penalty when the other FIRE-2 galaxies are excluded from training ($\sim 77\%$ in Training set~A) compared to when it is included ($\sim 66\%$ in Training set~B and C). Standardization eliminates this domain gap, stabilizing the scores at a consistent $\sim 63\%$ across all configurations, but the residual remains elevated relative to the other halos. Additionally, as seen in Table~\ref{tab:flow_moments}, \texttt{m12m} retains a 16\% error in the velocity standard deviation under stellar standardization, much higher than found in the other halos.

With the exception of this example of a highly-rotational dark disk, the other cross-training-set comparisons reveal robust transfer learning capabilities when combined with standardization. As an example, the unscaled baseline AUC for \texttt{halo685x09} varies considerably depending on which training set is used. Standardization smooths these fluctuations into $\sim 55\%$, confirming that our method captures invariant physical properties and enables the generative model to generalize robustly across distinct hydrodynamic implementations. 

\section{Discussion and Conclusion}~\label{sec:conclusion}

In this study, we demonstrated the efficacy of a standardization-based scaling strategy in revealing the universal nature of dark matter halos across diverse hydrodynamic simulations. We confirm that the kinematics of the DM bulk are universal and transferable within the Solar neighbourhood-like region, and through classification tasks we showed that structural differences between simulations can be minimized via statistical standardization to cover a larger domain of halo variation. This universality is physically expected, as the imprint of a galaxy's individual assembly history is progressively washed out of its present-day stellar chemodynamics over cosmic time~\citep{2026arXiv260504138N}, leaving a bulk phase-space structure that retains little memory of merger-specific details.

Second, we established that this universality can be practically recovered using metal-poor stars ([Fe/H] $<-2$) as dynamical tracers. We showed that scaling based on the stellar velocity dispersion successfully approximates the dark matter potential, providing a viable pathway to apply this framework to observational data from the Milky Way.

Finally, we demonstrated the utility of this universality for generative modelling. By training a MAF on a standardized multi-simulation ensemble, we achieved high-fidelity reconstruction of halo kinematics, 
 maintaining around 10\% accuracy within the the distribution bulk for most of the simulations. This implies that future research can leverage pooled datasets from multiple simulation suites to train robust, generalized models of the galactic phase-space, overcoming the sample variance limitations of individual simulations.

One halo, \texttt{m12m}, performed more poorly, with a recovery of the DM VPDF accurate to 25\% in $v_\phi$. This is because this simulations has a dark disc with $v_\phi\sim150$km/s, which is different from the rest of the simulated halos with 20 km/s rotation. We show that such fast-spinning feature in DM population can be revealed by the strong exsitu-stellar disk via the stellar motion measurement by Gaia. 

Beyond generative modeling, this universality benefits multi-galaxy kinematic studies. Pooling halos of widely different mass and size has traditionally biased such analyses~\citep{Bozorgnia:2017brl,Lilie:2025wkr}, as this scale variation dominates the combined sample and masks the kinematic structure of interest. By standardizing all halos, we remove variation and reveal the underlying shared phase-space form---allowing galaxies and simulations to be pooled and compared without scale-induced bias.

\section*{Data availability}
The \texttt{h277} galaxy data set is available in the N-Body Shop, at https://nbody.shop/data.html. \textbf{FIRE-2} galaxies are taken from FIRE-2 public data
release\citep{Wetzel:2022man}. The \textbf{Vintergatan-GM} galaxies are available under request. 

\section*{Acknowledgments}

JIR acknowledges support from STFC grants ST/Y002857/1 and ST/Y002865/1. MRB, DS, and SH are supported by the DOE under Award Number DOE-SC0010008. 

\appendix

\bibliography{reference}

\end{document}